\documentclass{PoS}

\title{Electroweak resonances in HEFT }

\ShortTitle{Electroweak resonances in HEFT }

\author{\speaker{Felipe J. Llanes-Estrada}\thanks{Backup speaker covering A. Dobado. 
Work supported by grants from MINECO FPA2011-27853-C02-01 and FPA2016-75654-C2-1-P, and carried out in the inspiring atmosphere of the theoretical physics department and UPARCOS. }\\
        E-mail: \email{fllanes@fis.ucm.es}}
\author{Rafael L. Delgado and Antonio Dobado\\
        Departamento de Fisica Teorica I, Plaza de las Ciencias 1, Fac. CC. Fisicas; Universidad Complutense de Madrid, 28040 Madrid, Spain\\
        E-mail: \email{rdelgadol@fis.ucm.es; dobado@fis.ucm.es}
}

\abstract{Due to the gap between the known 100 GeV scale and new physics if any, it is natural to employ an effective one-loop Lagrangian (HEFT) for the particles of the Electroweak Symmetry Breaking Sector ($W_L$, $Z_L$ and $h$). To describe any new particles and resonances that may be found at the LHC we employ its unitarized amplitudes, valid even in the presence of new strong interactions. We have assessed the systematics by comparing several such methods, and find that they give qualitatively similar results and succesfully produce unitary amplitudes in the nonperturbative regime. We are thus in a position to describe new physics in the 0.5 TeV-3 TeV (region of validity of our approximations: the effective theory and the equivalence theorem to substitute $W_L$, $Z_L$ by the Goldstone bosons of electroweak symmetry breaking). We have also computed the coupling of the EWSBS to the top-antitop and two-photon channels to describe resonances that decay through them or to study their photon-photon production, for example. The approach is universal and useful for many BSM theories at low energy.}

\FullConference{EPS-HEP 2017, European Physical Society conference on High Energy Physics\\
		5-12 July 2017\\
		Venice, Italy}

\begin{document}

The LHC collaborations ATLAS and CMS have discovered what looks like the Higgs boson of the Standard Model, which joins the longitudinal $W_L$ and $Z_L$ in the 100--GeV region Electroweak Symmetry Breaking Sector (EWSBS). But they have, as of this EPS-HEP conference, failed to report new physics. As the bounds on production cross sections tighten, and no resonances appear in the 1-2 TeV region~\cite{Aaboud:2017eta,Sirunyan:2017acf}, one may wonder how to proceed if, after all, new physics related to the EWSBS is beyond the energy/luminosity reach of the accelerator.  

\begin{figure}[h]
\centering
\includegraphics*[width=0.4\textwidth]{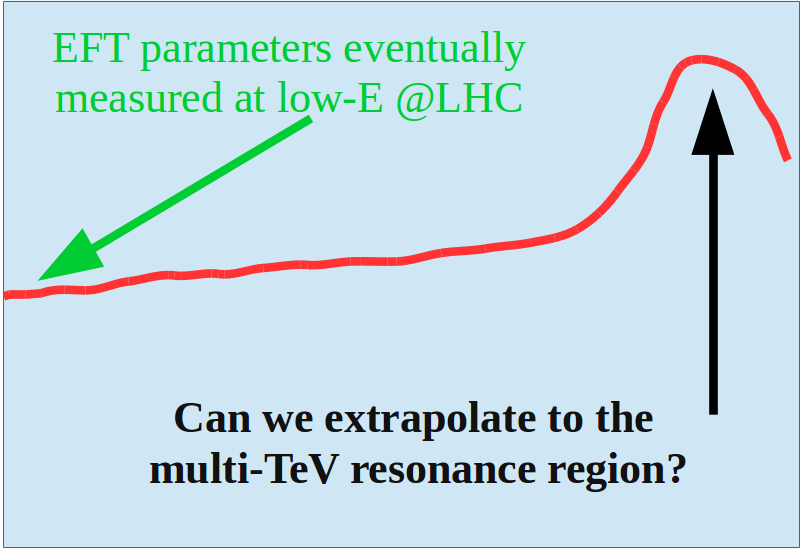}
\includegraphics*[width=0.54\textwidth]{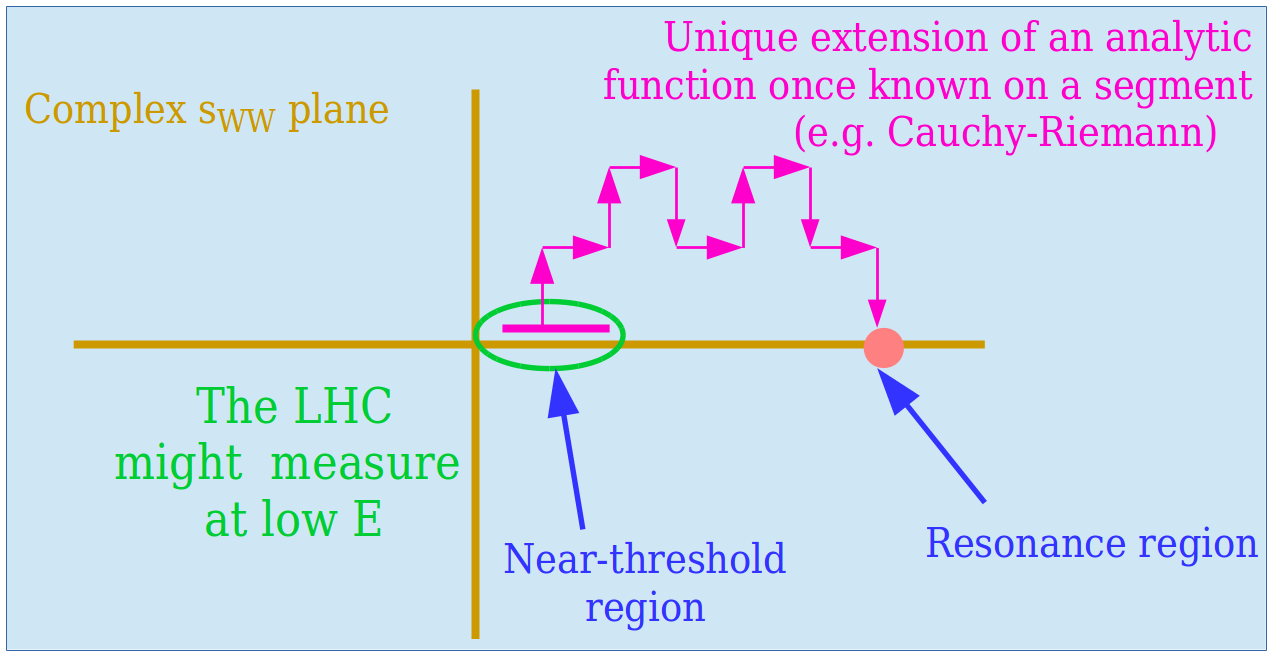}
\caption{\label{fig:taskathand}
{\bf Left plot}: the basic problem. If the LHC is able to measure a separation of the coefficients of the low-energy theory from the Standard Model, can we predict the presence of resonances at higher energy, perhaps outside the reach of the accelerator?
{\bf Right plot}: scattering partial waves of the low-energy particles $W_L$, $Z_L$, and $h$, are analytic in their upper-half complex $s$-plane due to causality. Thus if they are exactly known on a segment ({\it e.g.} at low energy), they can be extrapolated to the resonance region (and everywhere in the domain of analyticity). Thus, as a matter of principle, and given enough experimental precision, one can predict the presence of resonances of the EWSBS from low-energy measurements.}
\end{figure}

One way to proceed (common to lower energy experiments) is to carry out precision studies of the already known particles. A precise knowledge of the scattering partial waves of the $V_L$ and $h$ particles might reveal a separation from the Standard Model. In that case, can one predict whether a new resonance is within reach of the LHC or a successor machine? (see the sketch in figure~\ref{fig:taskathand}). 

Indeed, because of the analyticity of partial--wave amplitudes in the upper--half complex s--plane, one can in principle predict whether or not and where resonances can be found from low--energy measurements of those partial wave amplitudes. The devil is in the detail.

First, one needs a parametrization of the 100 GeV--energy region EWSBS. Two effective field theories extend the Standard Model~\cite{deFlorian:2016spz}. The first, SMEFT, by adding operators of higher canonical dimension, turning it into a nonrenormalizable effective theory with new coefficients that can be constrained by experiment. The Higgs boson is part of a doublet field with the SM potential.
The second theory which we pursue, HEFT or Higgs Effective Field Theory, is agnostic as to the specific nature of the Higgs boson and couples it in the most general way compatible with the $SU(2)\times SU(2)\to SU(2)_c$ symmetry breaking pattern, taking for the leading order the nonlinear formulation of chiral perturbation theory.
The Lagrangian of HEFT, neglecting the masses $M_W\sim M_Z\sim m_h\sim 100$ GeV, as appropriate to explore the TeV region where resonances might appear, is
\begin{eqnarray} \label{bosonLagrangian}
{\cal L}  =  \frac{1}{2}\left(1 +2 a \frac{h}{v} +b\left(\frac{h}{v}\right)^2\right)
\partial_\mu \omega^a
\partial^\mu \omega^b\left(\delta_{ab}+\frac{\omega^a\omega^b}{v^2}\right)   
\nonumber +\frac{1}{2}\partial_\mu h \partial^\mu h 
  +  \frac{4 a_4}{v^4}(\partial_\mu \omega^a\partial_\nu \omega^a)^2 \nonumber  \\
+\frac{4 a_5}{v^4}( \partial_\mu \omega^a\partial^\mu \omega^a)^2
+\frac{g}{v^4} (\partial_\mu h \partial^\mu h )^2  
  + \frac{2 d}{v^4} \partial_\mu h\partial^\mu h\partial_\nu \omega^a  \partial^\nu\omega^a
+\frac{2 e}{v^4} \partial_\mu h\partial^\nu h\partial^\mu \omega^a \partial_\nu\omega^a.
\end{eqnarray}

\begin{figure}[h]
\centering
 \raisebox{-0.5\height}{\includegraphics[width=0.3\textwidth]{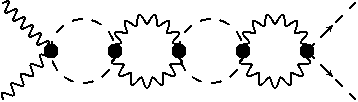}}\ \ \ \ \
 \raisebox{-0.5\height}{\includegraphics[width=0.5\textwidth]{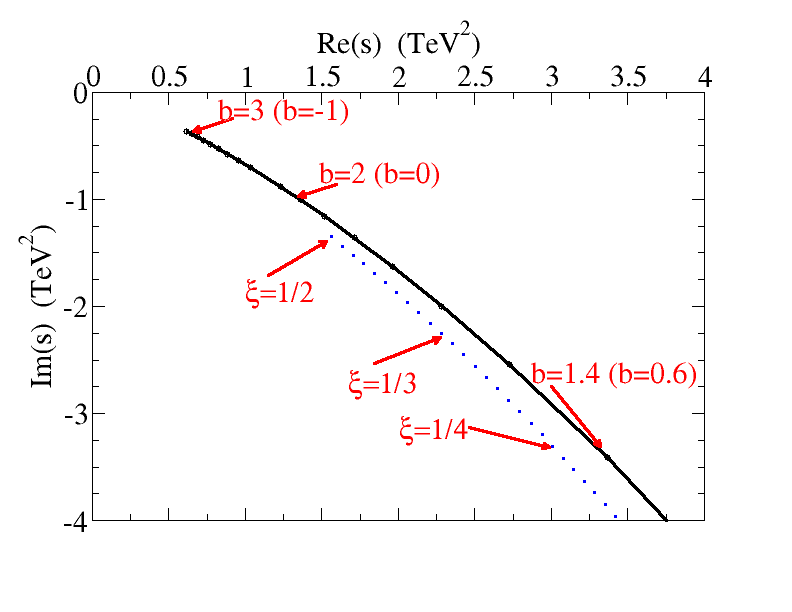}}
\caption{\label{fig:pinball}  {\bf Left}: typical Feynman-diagram interpretation of a pole in the $M(W_LW_L\to  hh)$ elastic amplitude, resonating between the two channels. Its experimental absence in the $W_LW_L$ forces a constraint on the $b$ parameter of the LO-HEFT. {\bf Right}: the pole motion (position of the resonance) in the complex $s$ plane. The SM value of $b=1$ sends it to infinite energy and removes it from the spectrum.}
\end{figure}

The LO parameters are, in the SM, $a=b=1$. A separation therefrom signals new physics. All other parameters are NLO and zero in the SM. The Equivalence Theorem allows us to use the Goldstone bosons $\omega$ \emph{en lieu des} longitudinal gauge bosons $W_L$, $Z_L$.
Amplitudes computed from this Lagrangian grow as $s^2$ up to logarithms, and thus the partial waves violate the unitarity bound $\arrowvert t\arrowvert \leq 1$. Indeed, the unitarity relation
for the partial waves $\arrowvert t \arrowvert^2= {\rm Im}(t)$ (for approximately massless particles, the phase space factor $\sigma$ can be taken as 1) is only satisfied to one less order in perturbation theory, for example ${\rm Im} (t^{(0)}+t^{(1)})= \arrowvert t^{0}\arrowvert^2$. 
Unitarisation methods solve this problem of perturbation theory representing the amplitudes in the resonance region by means of formulae that coincide at low energy with the EFT expansion but have better properties. A salient one is the Inverse Amplitude Method that we have described at length in the context of HEFT~\cite{Delgado:2015kxa} and that has the right analytic properties, satisfies elastic unitarity exactly, has been extended to coupled channels (feature that we exploit at length below) and matches the EFT amplitudes for small $s$. In addition, it is very compact,
\begin{equation}
t_{\rm IAM} = \frac{t^{(0)\ 2}}{t^{(0)}-t^{(1)}}\ .
\end{equation}
The structure of the denominator allows for poles of the amplitude on the second Riemann sheet of the complex $s$-plane, and this is the key to its use to parametrize resonances in terms of the HEFT.
A particularly interesting two-body channel is $hh$. Elastic $hh\to hh$ scattering is the textbook probe of the Higgs-sector potential. As of this conference, no signal has been reported in this Higgs-Higgs channel, which is actively searched for in $b\bar{b}b\bar{b}$ and $b\bar{b}\gamma\gamma$ among other final states, and the cross section is about 28 times that in the Standard Model~\cite{Morse:2017efg}. 
The LHC may not be sensitive enough to measure and explore this channel; but its $WW$ and $ZZ$ data can be used to constrain it together with the coupling parameter $b$. This application was highlighted by us~\cite{Delgado:2014dxa} and is illustrated in figure~\ref{fig:pinball}.

\begin{figure}
\centering
\includegraphics*[width=0.4\textwidth]{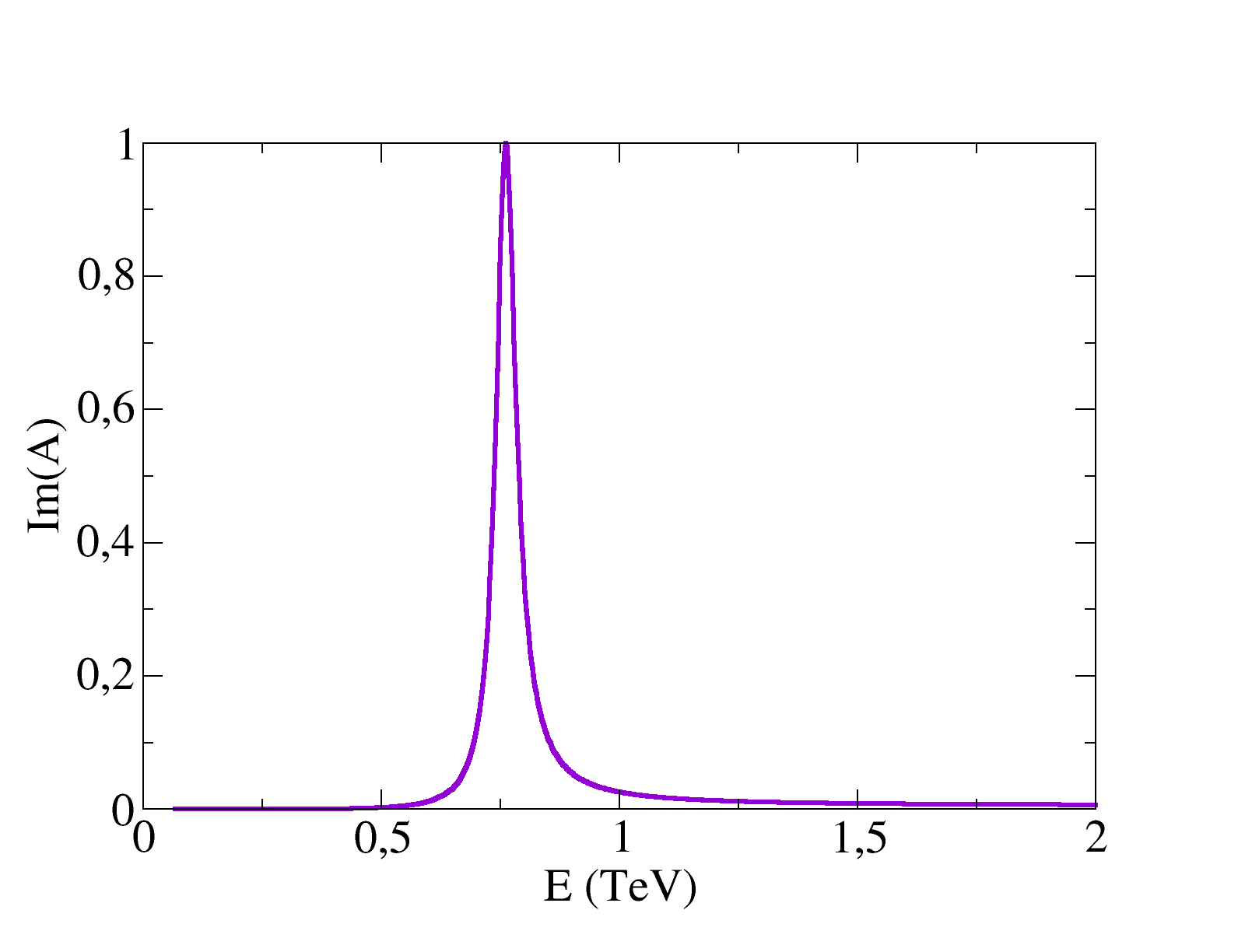}\ \
\includegraphics*[width=0.4\textwidth]{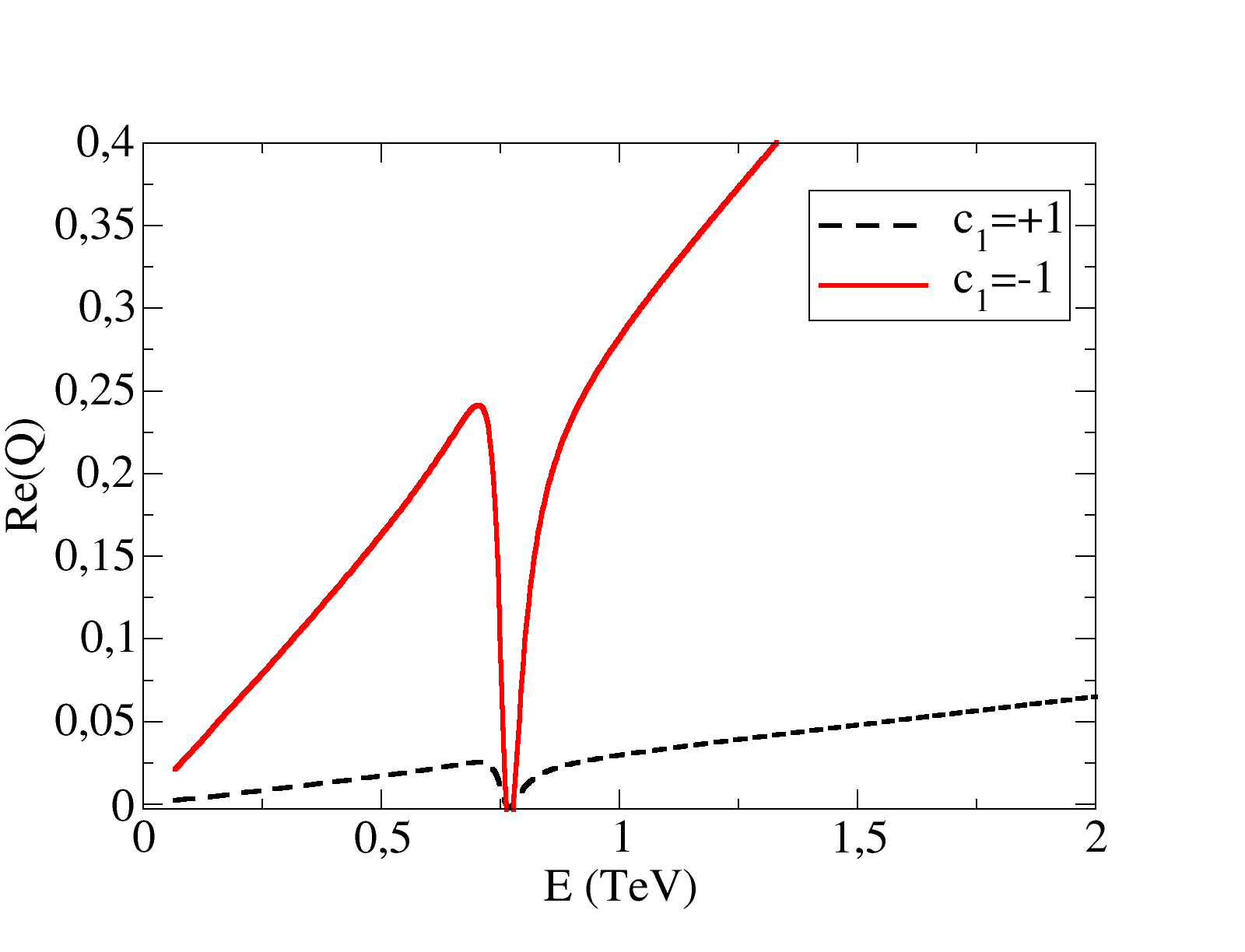}
\caption{\label{fig:top} {\bf Left}: a typical narrow resonance in the elastic $I=J=0$ $W_LW_L$ amplitude. {\bf Right}: The interference between the background $W_LW_L\to t\bar{t}$ and the resonant contribution can produce a dip in the $t\bar{t}$ spectrum.}
\end{figure}

LHC experiments can also try to find resonances of the EWSBS by looking into other channels that,
though normally not considered to have a strong effect on electroweak symmetry breaking, are sufficiently coupled that they can reflect the dynamics there. We have examined two cases,
The $t\bar{t}$ spectrum~\cite{Castillo:2016erh} and the $\gamma\gamma$ channel~\cite{Delgado:2016rtd}. The first is reproduced in figure~\ref{fig:top}. To produce it, we have 
expanded the basic Lagrangian by the addition of (as derived in~\cite{Castillo:2016erh}) appropriate couplings of the EWSBS and quark-antiquark sectors, 
\begin{eqnarray} \label{EFTLagrangianexpanded}
\mathcal{L}_{t\bar{t}} = -M_t\left(\!1\!+\!c_1\frac{h}{v}\!+\!c_2\frac{h^2}{v^2}\!\right)\bar{t}
\left\{\! \left(\!1-\frac{\omega^2}{2v^2}\right)\!  + \frac{i\sqrt{2}\omega^0}{v} \gamma^5  \!
\right\}t 
+ \frac{M_t}{v^4}t\bar{t} \left\{ g_t(\omega^i_{,\mu}\omega^{j\ ,\mu})  %
      +g'_t ( h_{,\mu}h^{,\mu}) \right\} 
\end{eqnarray}
where the LO  $c_1$, $c_2$ and  NLO $g_t$, $g'_t$ coefficients are \emph{a priori} unknown and control the intensity of the coupling between both sectors. Narrow resonances can be generated in the EWSBS sector by judicious choice of its parameters, and transfered to the $t\bar{t}$ spectrum via this Lagrangian, which need not be strongly coupled. Such resonances can appear as dips due to interference.

A similar reasoning leads to the effective Lagrangian coupling the EWSBS to the $\gamma\gamma$ channel,
\begin{eqnarray}
\mathcal{L}_{\gamma\gamma}= %
\frac{e^2 a_1}{2v^2}A_{\mu\nu}A^{\mu\nu}\left(v^2 - 4\omega^+\omega^-\right) %
 + \frac{2e(a_2-a_3)}{v^2}A_{\mu\nu}\left[%
         i\left(\partial^\nu\omega^+\partial^\mu\omega^- - \partial^\mu\omega^+\partial^\nu\omega^- \right) %
   \right. \\ \left. %
        +eA^\mu\left( \omega^+\partial^\nu\omega^- + \omega^-\partial^\nu\omega^+ \right)
        -eA^\nu\left( \omega^+\partial^\mu\omega^- + \omega^-\partial^\mu\omega^+ \right)
        \right] -\frac{c_{\gamma}}{2}\frac{h}{v}e^2 A_{\mu\nu} A^{\mu\nu} .
\end{eqnarray}
The parameters here are the electric charge $e$ (not to be confused with the NLO parameter in the HEFT above) and the NLO constants $a_i$ ($i=1\dots 3$) and $c_\gamma$~\cite{Delgado:2014jda}. The electromagnetic four-potential $A_\mu$ and field tensor $A_{\mu\nu}=\partial_\mu A_\nu-\partial_\nu A_\mu$ thus couple to the scalar sector of the theory with an intensity that is, at a minimum, as in scalar electrodynamics with the charge $e$, but that may include its corresponding anomalous couplings.

\begin{figure}\centering
\includegraphics[width=0.4\textwidth]{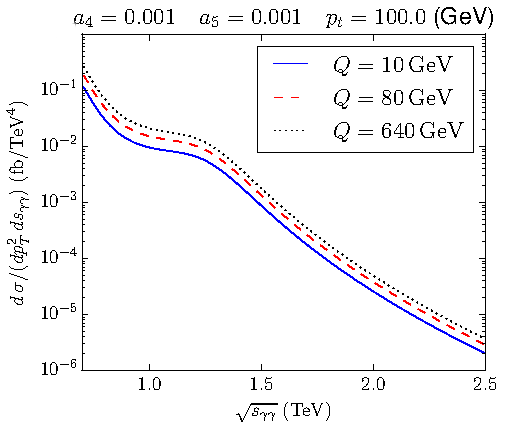}
\caption{\label{fig:fotones}
A narrow resonance in $W_LW_L$, generated by the NLO counterterms $a_4=10^{-3}$ and
$a_5=10^{-3}$ (while the LO parameters are $a=0.81$, $b\simeq a^2$ and the renormalization scale is $\mu=3$ TeV), is produced in 13 TeV collisions via intermediate $\gamma\gamma$ states. We plot the doubly differential cross section at $p_t=100$ GeV. The photon flux factors are taken from the LUX-QED package~\cite{Manohar:2016nzj}.}
\end{figure}

There are two obvious applications of this Lagrangian. The first is, as in the $t\bar{t}$ case, to describe resonances in the $\gamma\gamma$ final state. A second interesting one is to calculate production cross-sections in which the $W_LW_L$ pair (eventually $hh$ pair) arises from 
an intermediate $\gamma\gamma$ state (see figure~\ref{fig:fotones}).

This production mode can be of interest because the high energy protons at the LHC have a large classical EM field and the accelerator is thus partly a photon-photon collider, in the Weizs\"acker-Williams picture. Experimentally, the channel is clean because a large part of the cross section is elastic or quasielastic, so that at least one of the protons can be tagged forward and there is little additional in the central region.

In summary, HEFT methods extend the SM in the most general way, by coupling the newly found scalar $h$ boson to the longitudinal gauge bosons without prejudice as to the precise dynamical mechanism behind electroweak symmetry breaking.

Analyticity allows to use any constrains that the LHC can impose on this Lagrangian to predict the presence or absence of resonances at higher energies, even outside the reach of the accelerator. A competitive way of achieving this extension of low--energy amplitudes is the Inverse Amplitude Method, but we have also pursued the N/D method and the (improved, analytic) K-matrix method, as well as large-$N$ limit resummations and others.
Numerous applications are possible, and we have highlighted the couplings of $W_LW_L$ to three two--body channels, $hh$, $t\bar{t}$ and $\gamma\gamma$.


\end{document}